\documentclass[preprint,aps,showpacs,showkeys,dvips]{revtex4}

\usepackage{graphicx}
\usepackage{dcolumn}
\usepackage{bm}
\usepackage{subfigure}

\begin{document}

\title{Possible Single Resonant Production of the Fourth Generation Charged Leptons at $\gamma e$ Colliders}

\author{R. \c{C}ift\c{c}i}
\email{rena.ciftci@cern.ch}
\affiliation{Dept. of Eng. of Physics, Faculty of Eng., Ankara University, 06100 Tandogan,
Ankara, Turkey}

\author{ A. K. \c{C}ift\c{c}i}
\email{ciftci@science.ankara.edu.tr}
\affiliation{Physics Department, Faculty of Sciences, Ankara University, 06100 Tandogan,
Ankara, Turkey}

\author{S. Sultansoy}
\email{ssultansoy@etu.edu.tr}
\affiliation{Physics Section, Faculty of Sciences and Arts, TOBB University of Economics and Technology,
Ankara, Turkey}
\altaffiliation[Also at ]{Institute of Physics, Academy of Sciences, H. Cavid Avenue 33, 
Baku, Azerbaijan}
\date{\today}

\begin{abstract}
Single resonant productions of the fourth standard model generation charged lepton via anomalous interactions at $\gamma e$ colliders based on future linear $e^+ e^-$ colliders with $500$ GeV and $1$ TeV center of mass energies are studied. Signatures of $\gamma e\rightarrow \ell_4\rightarrow e\gamma $ and $\gamma e\rightarrow \ell_4\rightarrow eZ$ anomalous processes followed by the hadronic and leptonic decay of the $Z$ boson and corresponding standard model backgrounds are discussed. 
\end{abstract}
\maketitle

As the LHC run approaches, the interest on fourth standard model (SM) generation is increasing \cite{Holdom09}. Actually, existence of the fourth SM family follows from the standard model basics together with mass pattern of the third family fermions \cite{Fritzsch87,Datta93,Celikel95,Sultansoy07}. The new quarks (if they exist) will be copiously produced at the LHC \cite{Arik99,ATLAS99,Holdom06,Ozcan08}, whereas the observation of the new leptons is problematic due to rather small production cross section and large background. Obviously, the best place to investigate the new leptons will be  the linear $e^+ e^-$ colliders with sufficiently high center of mass energy. If the center of mass energy is not enough  for the pair production, single production can be considered. However, the single production of new charged lepton within SM seems not to be so promising \cite{Sher}.

Since the fourth family fermions are expected to be heavy, they can serve us as a window for new physics. The Ref. \cite{Fritzsch99} argues that due the large mass of the $t$ quark, its anomalous interactions should be enhanced compared to that of the lighter SM fermions. Obviously, this argumentation is even more valid for fourth SM family fermions.

It is well known that, the linear $e^+ e^-$ colliders provide opportunity to construct $\gamma e$ and $\gamma \gamma$ colliders by producing real high energy $\gamma$ beam through Compton backscattering of laser photons from high energy lepton beam (see \cite{Telnov08} and references therein).

In this paper, we consider single anomalous production of the fourth SM family lepton at future $\gamma e$ colliders. The processes $\gamma e\rightarrow \ell_4\rightarrow e\gamma $ and $\gamma e\rightarrow \ell_4\rightarrow eZ$ ($Z\rightarrow \ell^+ \ell^-$, $jet jet$) are studied.

\begin{figure}[b]
\vspace{-0.2cm}
\subfigure[]{\includegraphics[width=6.25cm]{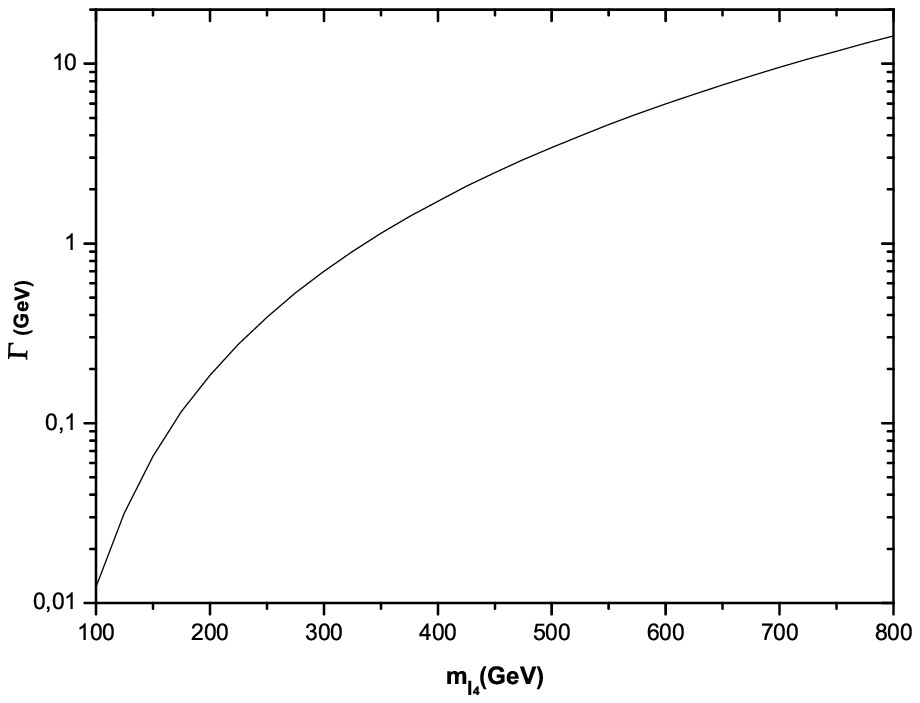}} \ \ \ \ \ \ \ \ \ \ 
\subfigure[]{\includegraphics[width=6.25cm]{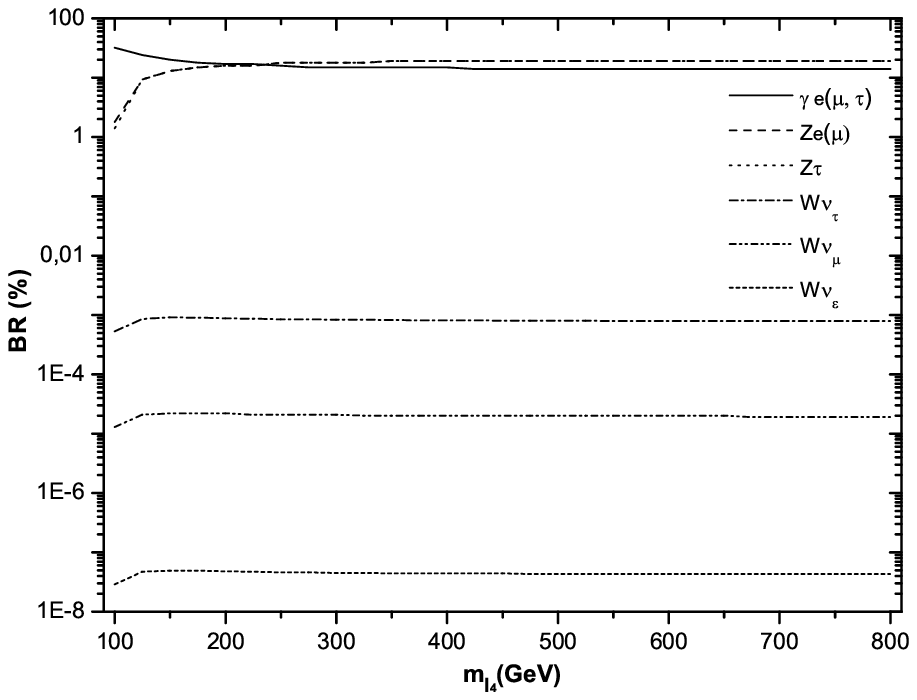}}
\vspace{-0.2cm}
\caption{(a) The total decay width of the fourth generation charged lepton and (b) the branching ratios (\%) depending on the mass of the charged lepton. \protect\label{fig1}}
\end{figure}  

\begin{figure}[b]
\vspace{-0.2cm}
\subfigure[]{\includegraphics[width=6.25cm]{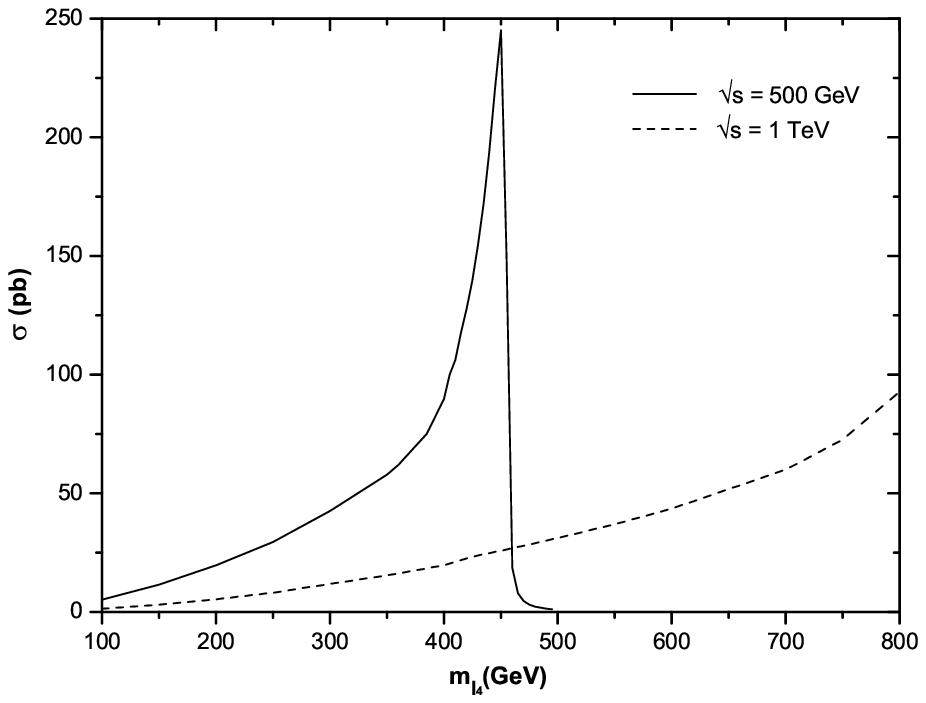}} \ \ \ \ \ \ \ \ \ \ 
\subfigure[]{\includegraphics[width=6.25cm]{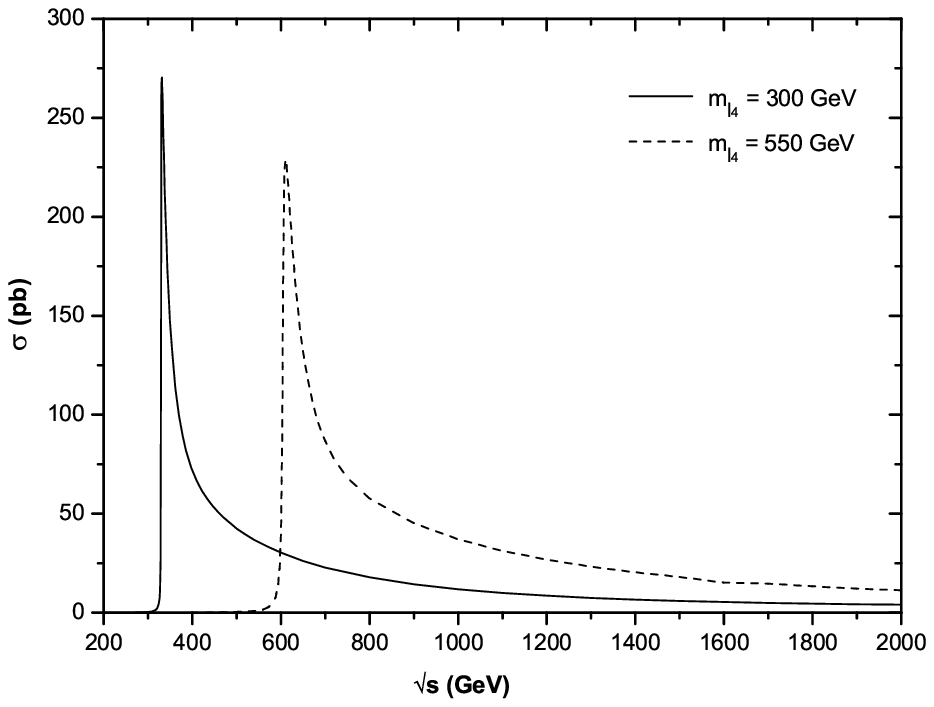}}
\vspace{-0.2cm}
\caption{The production cross sections of the fourth generation charged lepton at $\gamma e$ colliders based on $e^+ e^-$ machines (a) as a function of the lepton mass with the fixed center of mass energies of the collider and (b) as a function of center of mass energy of the collider with the fixed masses of the charged lepton. \protect\label{fig2}}
\end{figure} 

\begin{figure}
\vspace{-0.2cm}
\subfigure[]{\includegraphics[angle=-90,width=5cm]{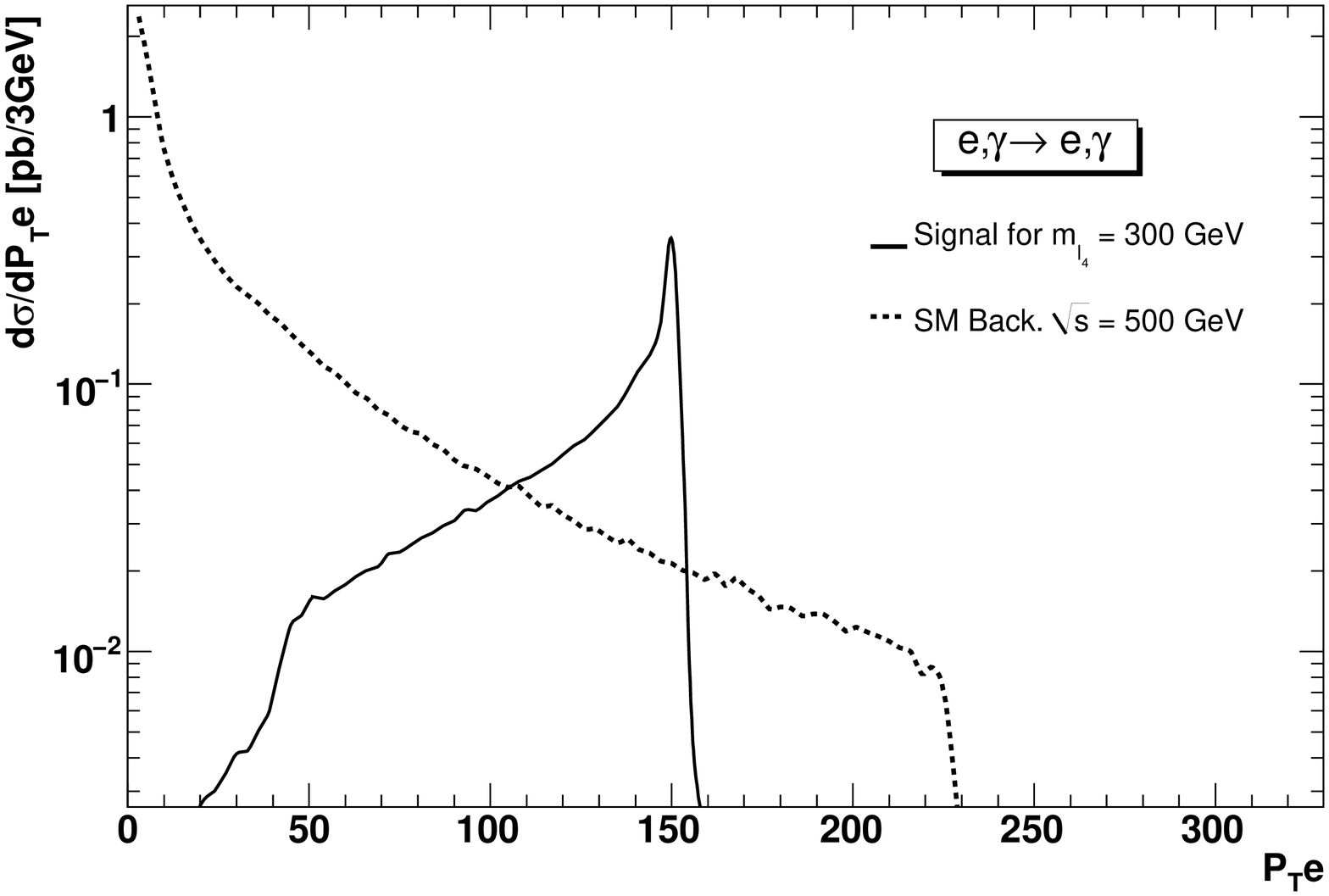}} \ \ \ \ 
\subfigure[]{\includegraphics[angle=-90,width=5cm]{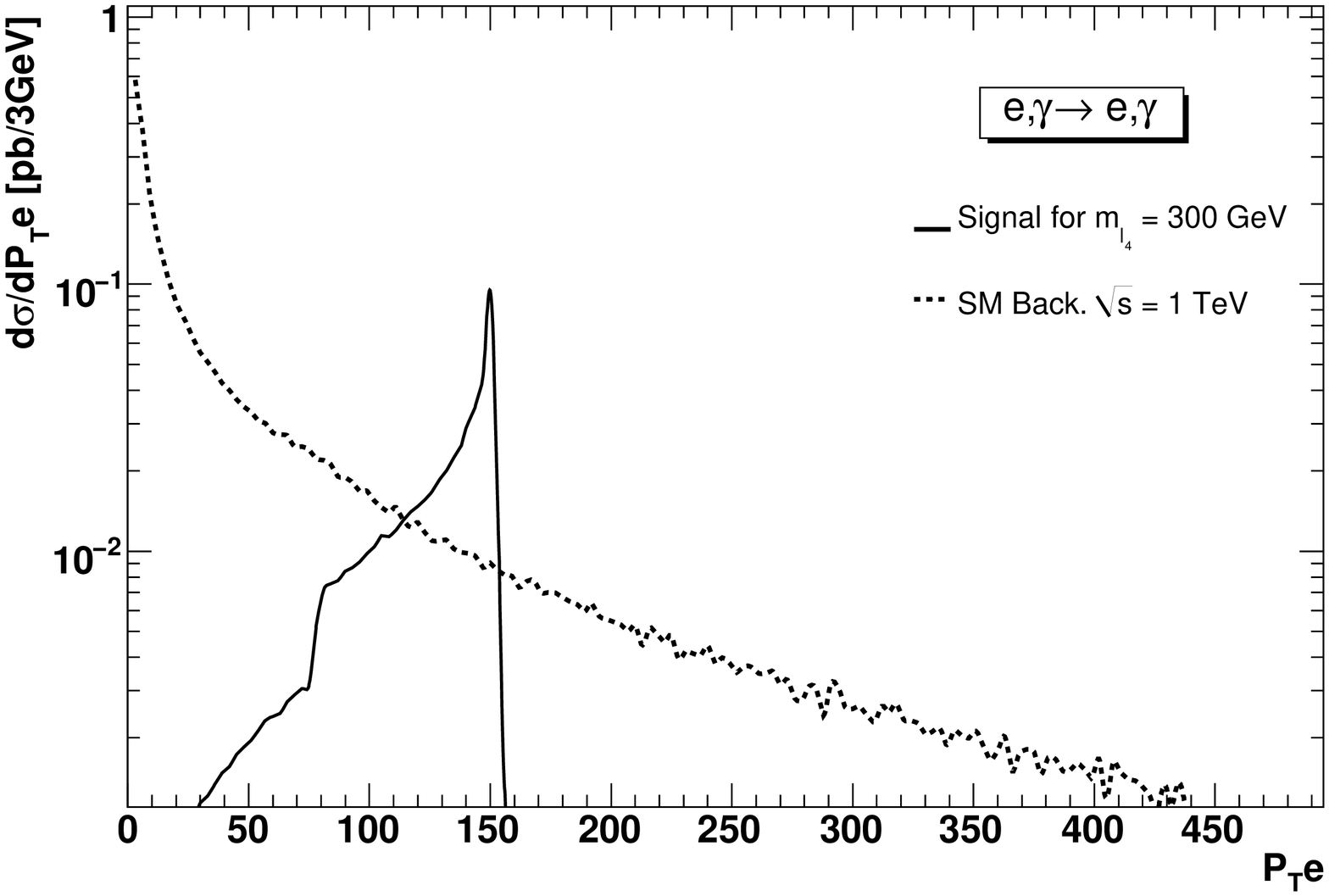}} \ \ \ \ 
\subfigure[]{\includegraphics[angle=-90,width=5cm]{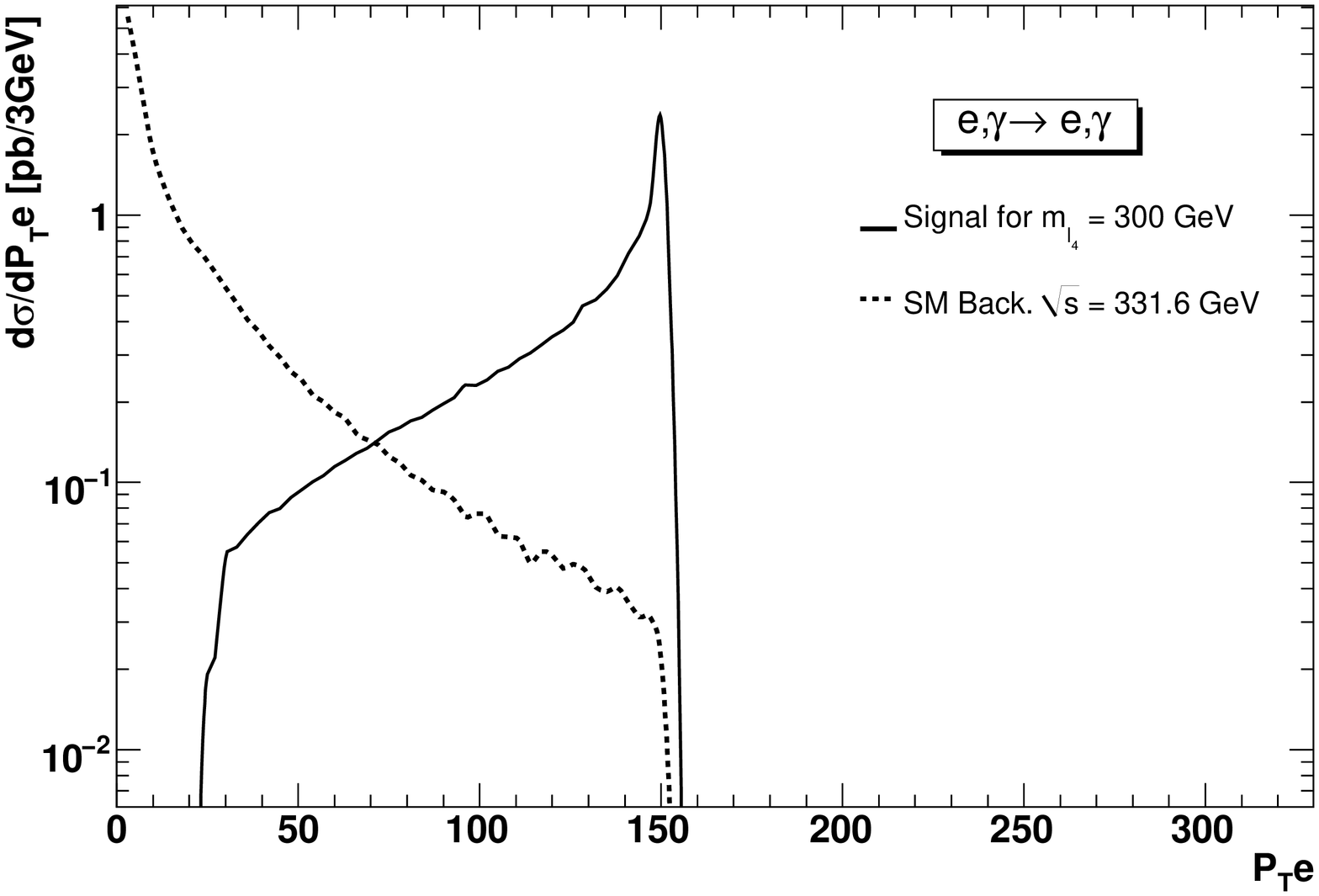}}
\vspace{-0.2cm}
\caption{The $P_{T}$ distributions for $\gamma e \rightarrow \ell_4\rightarrow e\gamma $ signal with 300 GeV lepton mass and corresponding SM background (a) at $\sqrt{s}=500$ GeV, (b) at $\sqrt{s}=1$ TeV and (c) at $\sqrt{s}=331.6$ GeV (at the resonance). \protect\label{fig3}}
\end{figure}

\begin{figure}[b]
\vspace{-0.2cm}
\subfigure[]{\includegraphics[angle=-90,width=5cm]{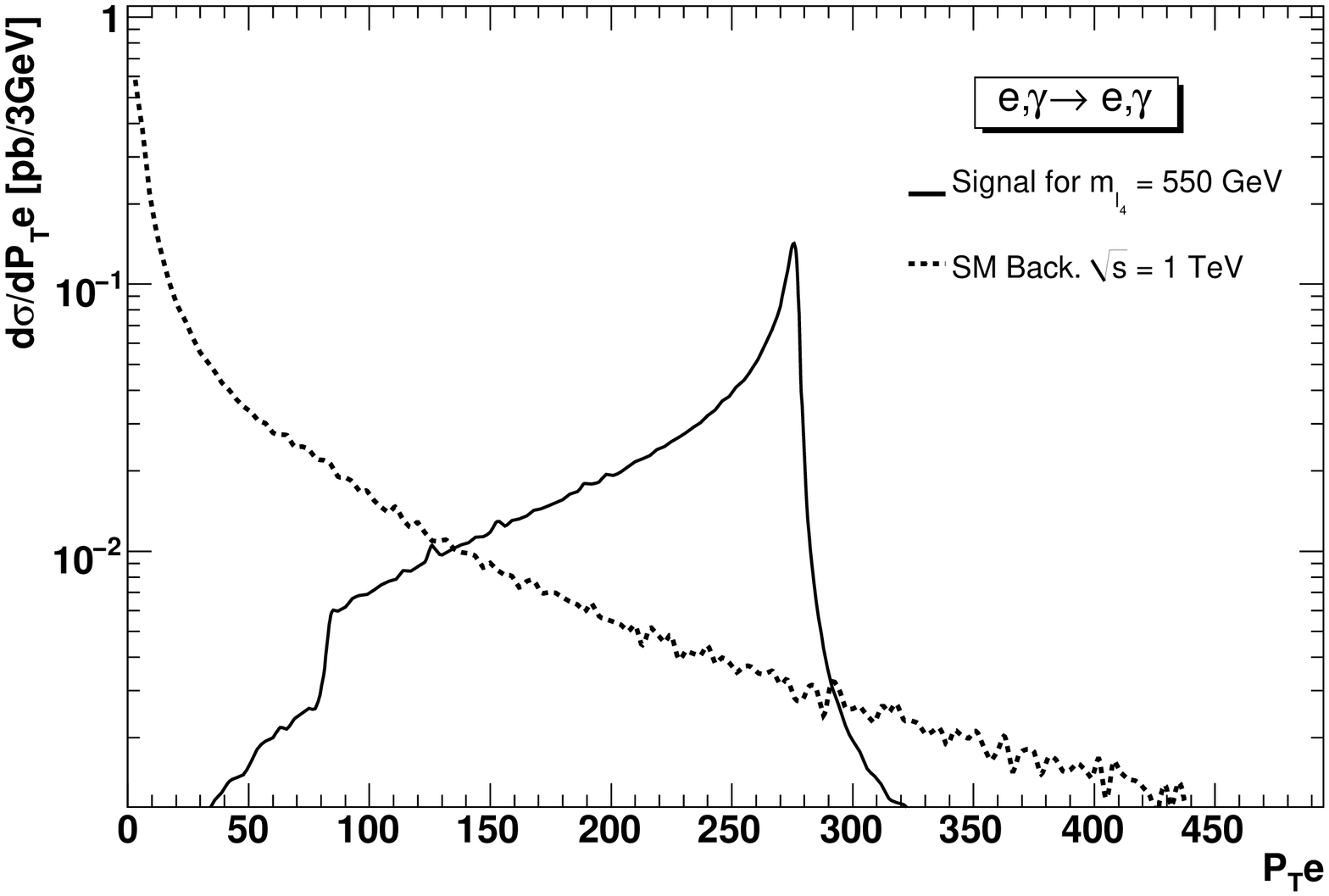}} \ \ \ \ \ \ \ \ \ \
\subfigure[]{\includegraphics[angle=-90,width=5cm]{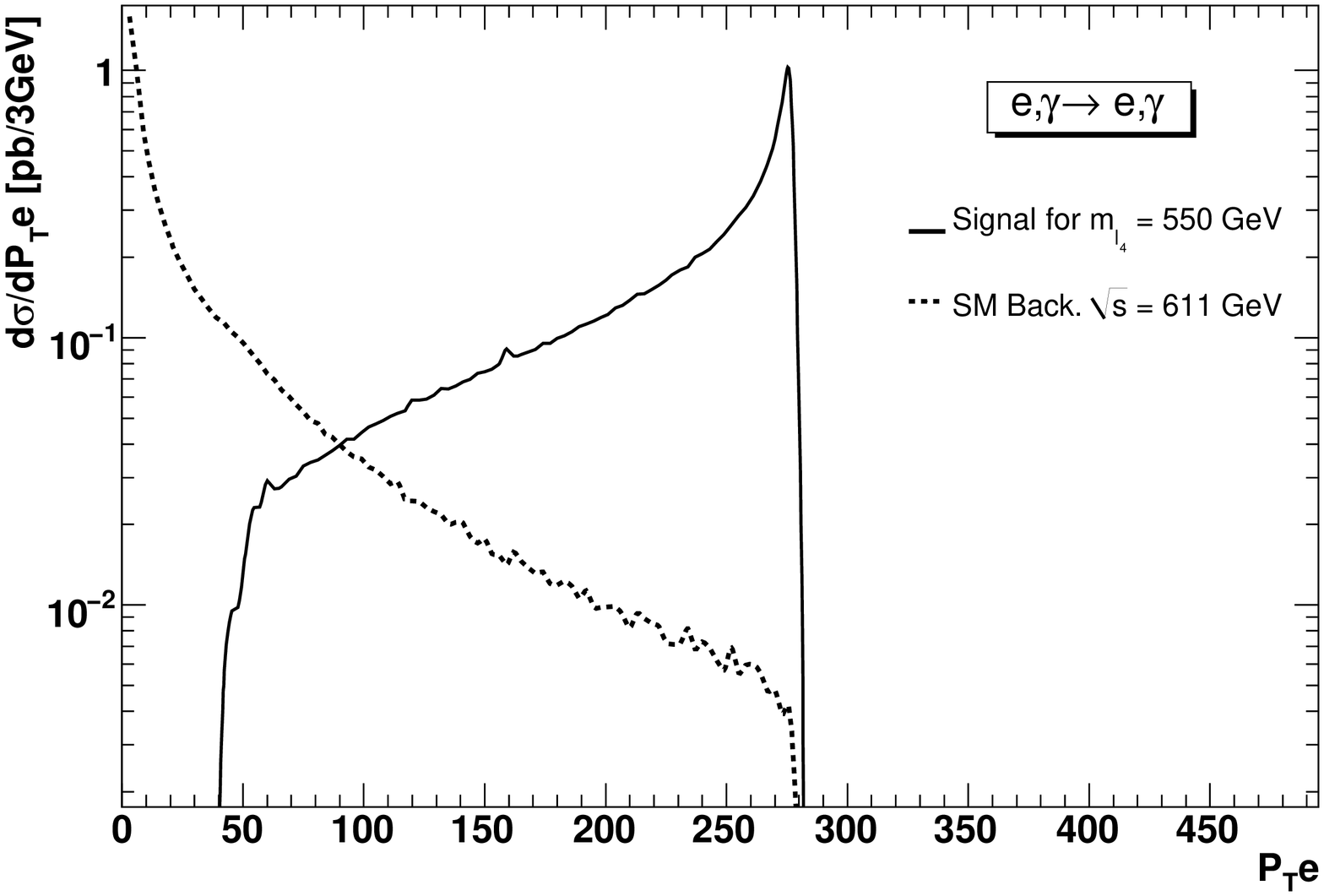}}
\vspace{-0.2cm}
\caption{The $P_{T}$ distributions for $\gamma e\rightarrow \ell_4\rightarrow e\gamma $ signal with 550 GeV lepton mass and corresponding SM background (a) at $\sqrt{s}=1$ TeV and (b) at $\sqrt{s}=611$ GeV (at the resonance).\protect\label{fig4}}
\end{figure}

The anomalous interactions, may rise as prescribed in Ref. \cite{Fritzsch99}, cause the flavor changing neutral currents (FCNC). The effective Lagrangian for the magnetic type FCNC of the fourth generation charged lepton can be written in a similar manner with \cite{Rizzo97}
\begin{equation}
\vspace{-0.2cm}
L=\left(\frac{\kappa^{\ell_{i}}_\gamma}{\Lambda}\right)e_{\ell}g_{e}\overline{\ell_{4}}\sigma_{\mu\nu}\ell_{i}F^{\mu\nu}+\left(\frac{\kappa^{\ell_{i}}_Z}{2\Lambda}\right)g_{Z}\overline{\ell_{4}}\sigma_{\mu\nu}\ell_{i}Z^{\mu\nu}+h.c. \hspace{2mm} \\
\vspace{-0.2cm}
\end{equation}\\
where $i=1,2,3$ correspond to the SM generation numbers; $\kappa^{\ell}_{\gamma,Z}$ are the anomalous couplings for the neutral currents with a photon and a Z boson, respectively (in this study $\kappa^{\ell_{1}}_{\gamma,Z}=\kappa^{\ell_{2}}_{\gamma,Z}=\kappa^{\ell_{3}}_{\gamma,Z}=\kappa^{\ell}_{\gamma,Z}$ is taken for simplicity). $\Lambda$ is the cutoff scale for the new physics and $e_\ell$ is the charge of leptons; $g_{e}$ and $g_{Z}$ are the electroweak coupling constants; $g_{e}= \sqrt{4\pi \alpha_{em}}$, $g_{Z}=g_{e}/cos\theta_{W}sin\theta_{W}$, where $\theta_{W}$ is the Weinberg angle. In the above equation, $\sigma_{\mu\nu} = i(\gamma_{\mu}\gamma_{\nu}-\gamma_{\nu}\gamma_{\mu})/2$. $F^{\mu\nu}$ and $Z^{\mu\nu}$ are field strength tensors of the photon and Z boson, respectively.

We have implemented the new interaction vertices into the CompHEP \cite{Boos04} for computations. Naturally, the anomalous interactions will introduce the additional decay channels of the fourth generation charged lepton. While calculating the SM decay contributions, we have used values of the Pontecorvo-Maki-Nakawaga-Sakata (PMNS) mixings given in Ref. \cite{Ciftci05}. The total decay width of the fourth generation charged lepton and the relative branching ratios are presented in Fig. 1, where $(\kappa_{\gamma}/\Lambda)=(\kappa_{Z}/\Lambda)=(\kappa/\Lambda)=1$ TeV$^{-1}$ is taken. We have used this value at the rest of our calculations. One can rescale our results keeping in mind that anomalous decay widths are proportional to $(\kappa/\Lambda)^2$.

The anomalous single production cross sections of the fourth SM generation charged lepton at $\gamma e$ colliders based on future linear $e^+ e^-$ colliders with $\sqrt{s}=500$ GeV and $1$ TeV are given in Fig. 2a and 2b, respectively. As seen from Fig. 2a the maximum production cross section of the fourth generation charged lepton at $\gamma e$ colliders based on $e^+ e^-$ machines with $\sqrt{s}=500$ GeV is $245$ pb for $450.5$ GeV lepton mass. This mass value can be easily understood from kinematics of the $\gamma e$ collider. Indeed, high energy photons can acquire $81 \%$ of electron energy at maximum \cite{Telnov08}. Therefore,  $(\sqrt{s_{\gamma e}})_{max} \cong 0.9 \sqrt{s_{e^+ e^-}}$.  Eventhough the maximum production cross section for $\sqrt{s}=1$ TeV option is reached at $901$ GeV mass value of the lepton, it is plotted in the figure until $800$ GeV. The last value is close to the upper limit on heavy fermion masses, which follows from partial-wave unitary at high energies \cite{Chanowitz79}. 

In principle, adjusting the center of mass energy of $\gamma e$ collider is possible and one can scan $\sqrt{s}$ to find resonance peak.  It is possible to decrease the maximum energy of photons by changing the angle of laser with respect to the electron beam. Therefore, one can optimize the center of mass energy of the collider in appropriate manner. As seen from Fig. 2b $\sqrt{s}=331.6$ GeV and $611$ GeV are ideal center of mass energies to produce fourth generation charged lepton with $300$ and $550$ GeV masses, respectively.

\begin{table}
\caption{The signal and SM background cross sections for anomalous processes at $\gamma e$ colliders based on $e^+ e^-$ machines with $\sqrt{s}=500$ GeV (at $(\kappa/\Lambda)$ = 1TeV$^{-1}$). Cut selection criteria are following: Selection 1 is $\left|\eta_{e,\ell,j}\right|\:<\:2.5$ and Selection 2 is $\left|\eta_{e,\ell,j}\right|\:<\:2.5$, $P^{e}_{T}\:>\:100$ GeV and $\Delta R_{jj}\:>\:0.4$.}  
\begin{ruledtabular}
\begin{tabular}{ccccccc}
 &\multicolumn{6}{c} {Signal cross sections (pb)} \\
 & \multicolumn{2}{c}{$\gamma e\rightarrow e\gamma $}&\multicolumn{2}{c}{$\gamma e\rightarrow e \ell \ell$}  & \multicolumn{2}{c}{$\gamma e\rightarrow e j j$}  \\ 
 \cline{2-3} \cline{4-5}\cline{6-7}
 $m_{\ell_{4}}$ (GeV)&Selection 1&Selection 2&Selection 1&Selection 2&Selection 1&Selection 2\\
\hline
300 & 6.3 & 4.8 & 0.50 & 0.35 & 5.0 & 3.5\\
350 & 8.5 & 7.1 & 0.72 & 0.58 & 7.2 & 5.9\\
400 & 13 & 12 & 1.2 & 1.0 & 12 & 10\\
425 & 19 & 17 & 1.7 & 1.5 & 17 & 15\\
450 & 34 & 31 & 3.0& 2.7 & 30 & 27\\
\hline
SM Backg. (pb)&32 & 2.6 & 2.5& 0.26 & 3.5 & 0.61\\
\end{tabular}
\end{ruledtabular}
\end{table}

\begin{table}
\caption{The same as Table I but for $\sqrt{s}=1$ TeV.}  
\begin{ruledtabular}
\begin{tabular}{ccccccc}
 &\multicolumn{6}{c} {Signal cross sections (pb)} \\
 & \multicolumn{2}{c}{$\gamma e\rightarrow e\gamma$ }&\multicolumn{2}{c}{$\gamma e\rightarrow e \ell \ell$}  & \multicolumn{2}{c}{$\gamma e\rightarrow e j j$}  \\ 
 \cline{2-3} \cline{4-5}\cline{6-7}
 $m_{\ell_{4}}$ (GeV)&Selection 1&Selection 2&Selection 1&Selection 2&Selection 1&Selection 2\\
\hline
300 & 1.7 & 1.3 & 0.12 & 0.090 & 1.2 & 0.94\\
400 & 2.8 & 2.6 & 0.23 & 0.21 & 2.3 & 2.2\\
500 & 4.3 & 4.0 & 0.37 & 0.36 & 3.9 & 3.6\\
550 & 5.1 & 4.8 & 0.45 & 0.44 & 4.6 & 4.4\\
600 & 6.0 & 5.7 & 0.54 & 0.53 & 5.5 & 5.3\\
700 & 8.3 & 8.1 & 0.77 & 0.76 & 8.0 & 7.6\\
800 & 13 & 13 & 1.2& 1.2 & 12 & 12\\
\hline
SM Backg. (pb)& 9.1 & 1.6 & 0.64 & 0.34 & 0.93 & 0.54\\
\end{tabular}
\end{ruledtabular}
\end{table}

\begin{table}[b]
\caption{The same as Table I but for ideal center of mass energies for $m_{\ell_4}=300$ and $550$ GeV.}  
\begin{ruledtabular}
\begin{tabular}{cccccccc}
 & &\multicolumn{6}{c} {Signal cross sections (pb)} \\
 & & \multicolumn{2}{c}{$\gamma e\rightarrow e\gamma$ }&\multicolumn{2}{c}{$\gamma e\rightarrow e \ell \ell$}  & \multicolumn{2}{c}{$\gamma e\rightarrow e j j$}  \\ 
 \cline{3-4} \cline{5-6}\cline{7-8}
 $m_{\ell_{4}}$ (GeV)& $\sqrt{s_{ee}}$ (GeV) &Selection 1&Selection 2&Selection 1&Selection 2&Selection 1&Selection 2\\
\hline
300 & 331.6 & 40& 30 & 3.2 & 2.2 & 33 & 22\\
Backg. (pb)& & 66 & 2.4 & 5.5 & 0.14 & 3.3 & 0.37\\
\hline
550 & 611 & 32 & 30 & 2.9 & 2.8 & 29 & 28\\
Backg. (pb)& & 22& 2.4 & 1.7 & 0.30  & 2.6 & 0.61\\
\end{tabular}
\end{ruledtabular}
\end{table}

Below $\gamma e\rightarrow \ell_4\rightarrow e\gamma $ and $\gamma e\rightarrow \ell_4\rightarrow eZ$ signal processes followed by the hadronic and leptonic decay of the $Z$ boson ($j=u, d, s, c, b$ and $\ell = e, \mu$) as well as their SM backgrounds are considered. Some kinematic cuts have been applied in order to suppress the SM background. First, a generic $\left|\eta_{e,\ell,j}\right|\:<\:2.5$ cut is chosen, where $\eta$ denotes the pseudorapidity. After the $\eta$ cut we have plotted the $P_{T}$ distributions of the electron for $\gamma e\rightarrow \ell_4\rightarrow e\gamma $ and corresponding background processes for $300$ GeV lepton mass in Figs. 3 and for $550$ GeV lepton mass in Figs. 4 to determine the optimum $P_{T}$ cut value. It is seen that $P^{e}_{T}>100$ GeV removes the most of the background while preserves the most of the signal events.  Similar statement is valid for the remaining processes (we do not give corresponding figures to save the space). The computed signal and background cross sections for processes under consideration are presented in Tables I and II for $\sqrt{s}=500$ GeV and $1$ TeV options, respectively. Results of the similar analysis for ideal center of mass energy are presented in Table III. The advantage of the tuning of $\sqrt{s}$ at $\gamma e$ colliders is seen from the comparison of Table III with Tables I and II. 

It is clear that with $\kappa/\Lambda=1$ TeV$^{-1}$ one can discover the fourth generation charged lepton until the mass within kinematical limit of the collider. In order to obtain achievable values of the anomalous coupling strength, we require statistical significance (SS) greater than 3 and at least 5 events per working year ($10^7$ s) as observation criteria. SS values are evaluated from \cite{CMS}
\begin{equation}
\vspace{-0.2cm}
SS=\sqrt{2\times L_{\gamma e}[(\sigma_{s}+\sigma_{b})ln(1+\frac{\sigma_{s}}{\sigma_{b}})-\sigma_{s}]}
\end{equation}
where $L_{\gamma e}$ is the integrated luminosity of the $\gamma e$ collider, which is taken as $65 \%$ of $100$ fb$^{-1}$ for $\sqrt{s_{ee}}=500$ GeV and $65 \%$ of $300$ fb$^{-1}$ for $\sqrt{s_{ee}}=1$ TeV.

\begin{figure}[b]
\vspace{-0.2cm}
\subfigure[]{\includegraphics[width=6cm]{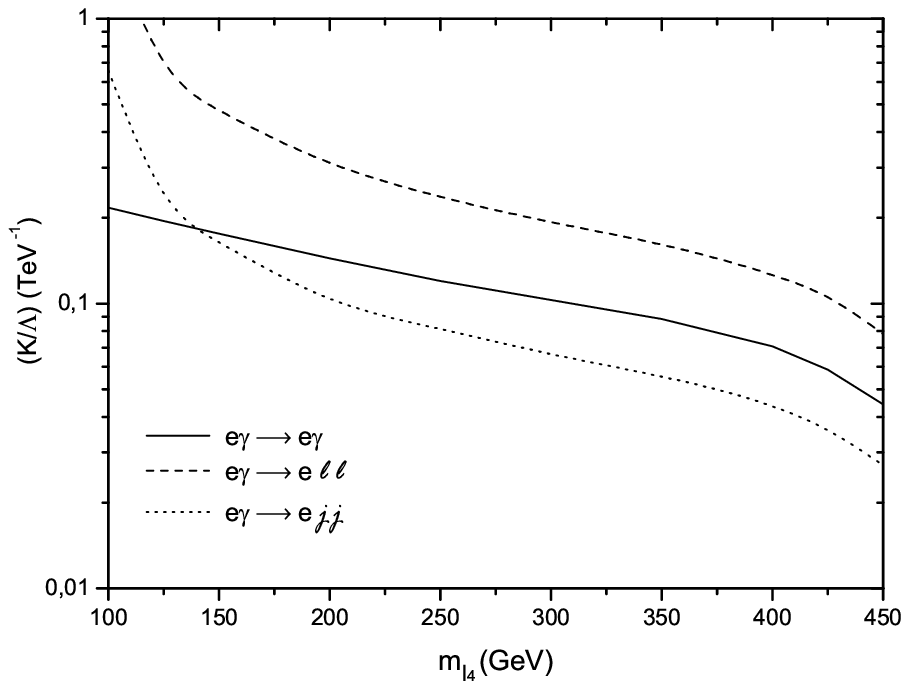}} \ \ \ \ \ \ \ \ \ \  
\subfigure[]{\includegraphics[width=6cm]{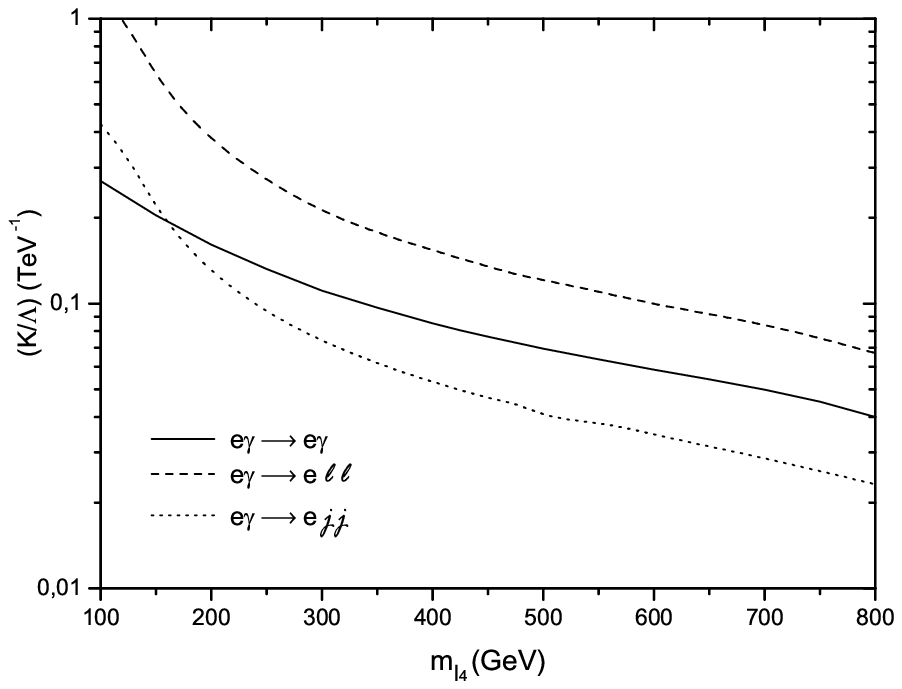}} 
\subfigure[]{\includegraphics[width=5cm]{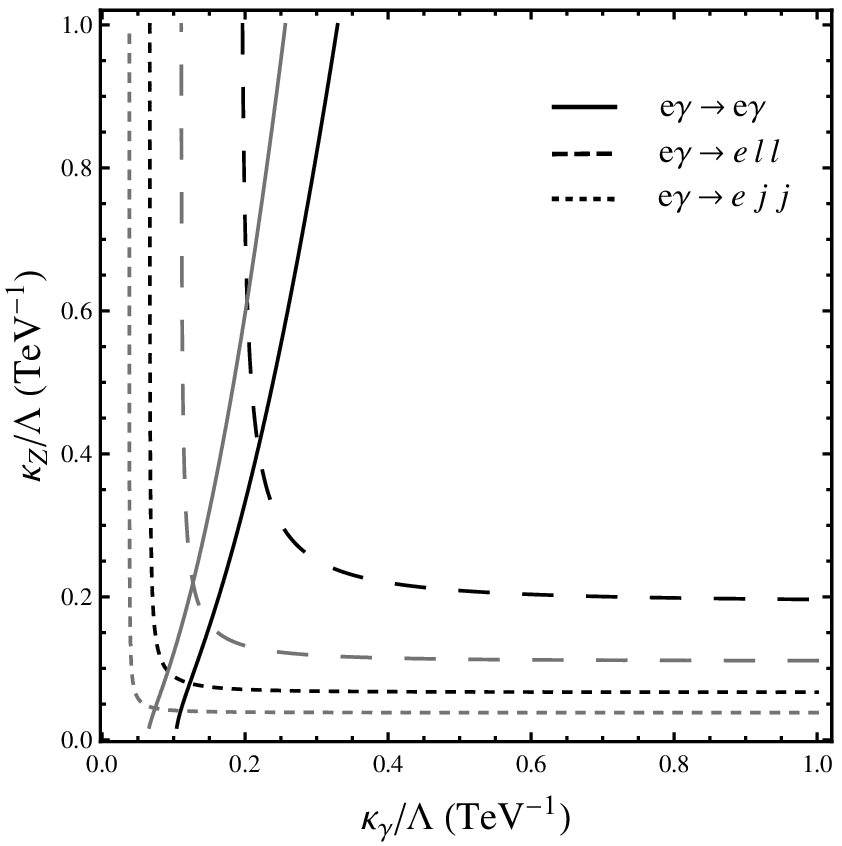}}
\vspace{-0.2cm}
\caption{The achievable values of the anomalous coupling strength at $\gamma e$ colliders based on $e^+ e^-$ machines with (a) $\sqrt{s}=500$ GeV and (b) $\sqrt{s}=1$ TeV as a function of the charged lepton mass; (c) the reachable values of anomalous photon and Z couplings for $m_{\ell_{4}}=300$ GeV at $\sqrt{s}=500$ GeV with $L_{int}=100$ fb$^{-1}$ (black lines) and for $m_{\ell_{4}}=550$ GeV at $\sqrt{s}=1$ TeV with $L_{int}=300$ fb$^{-1}$ (grey lines). Cut selection 1 is used. \protect\label{fig5}}
\end{figure}

\begin{figure}[b]
\vspace{-0.2cm}
\subfigure[]{\includegraphics[width=6cm]{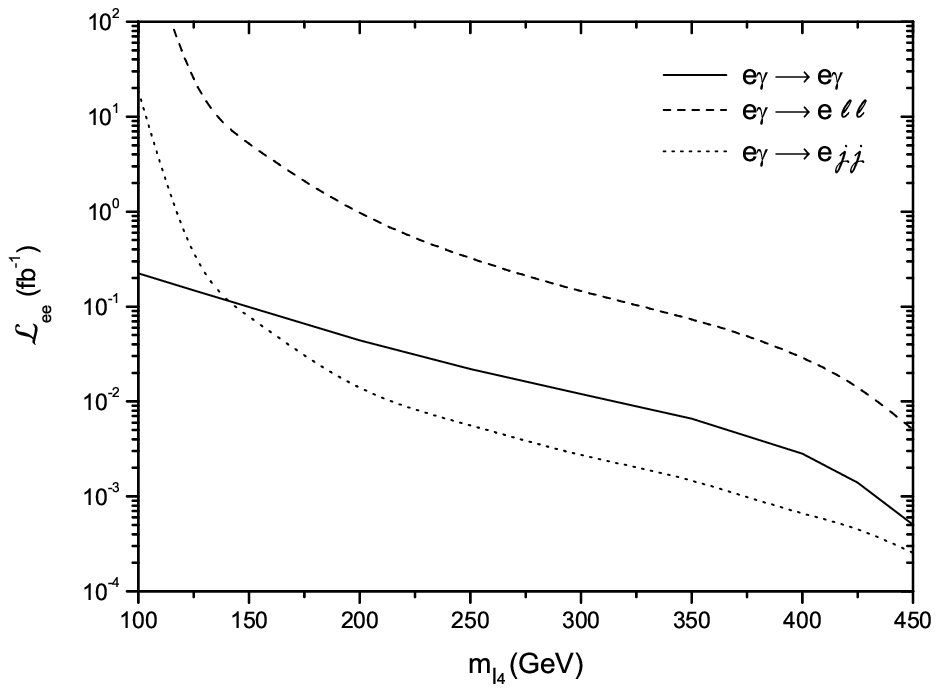}} \ \ \ \ \ \ \ \ \ \ 
\subfigure[]{\includegraphics[width=6cm]{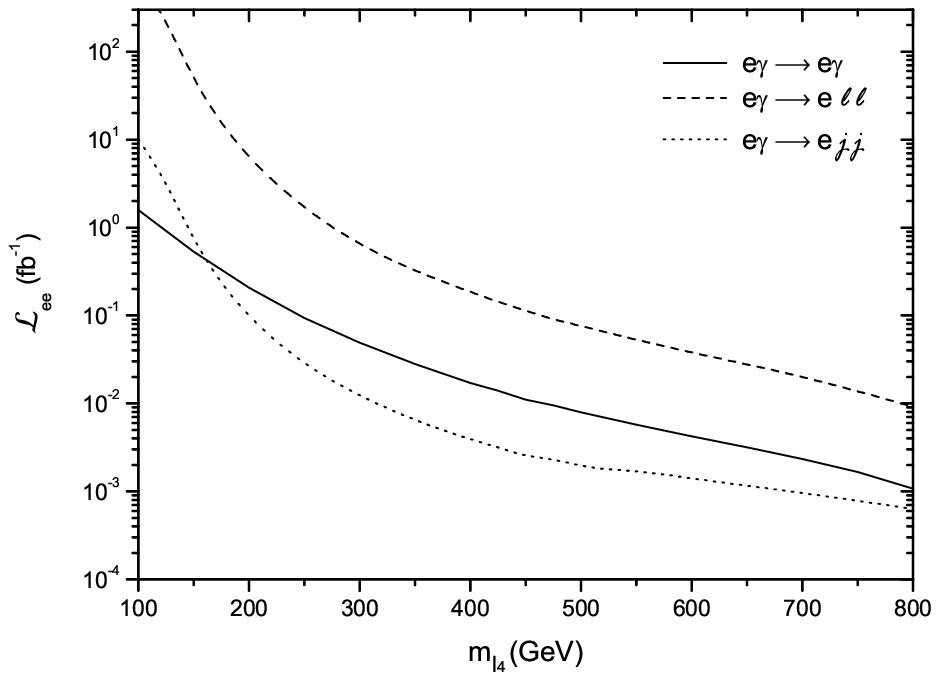}}
\subfigure[]{\includegraphics[width=6cm]{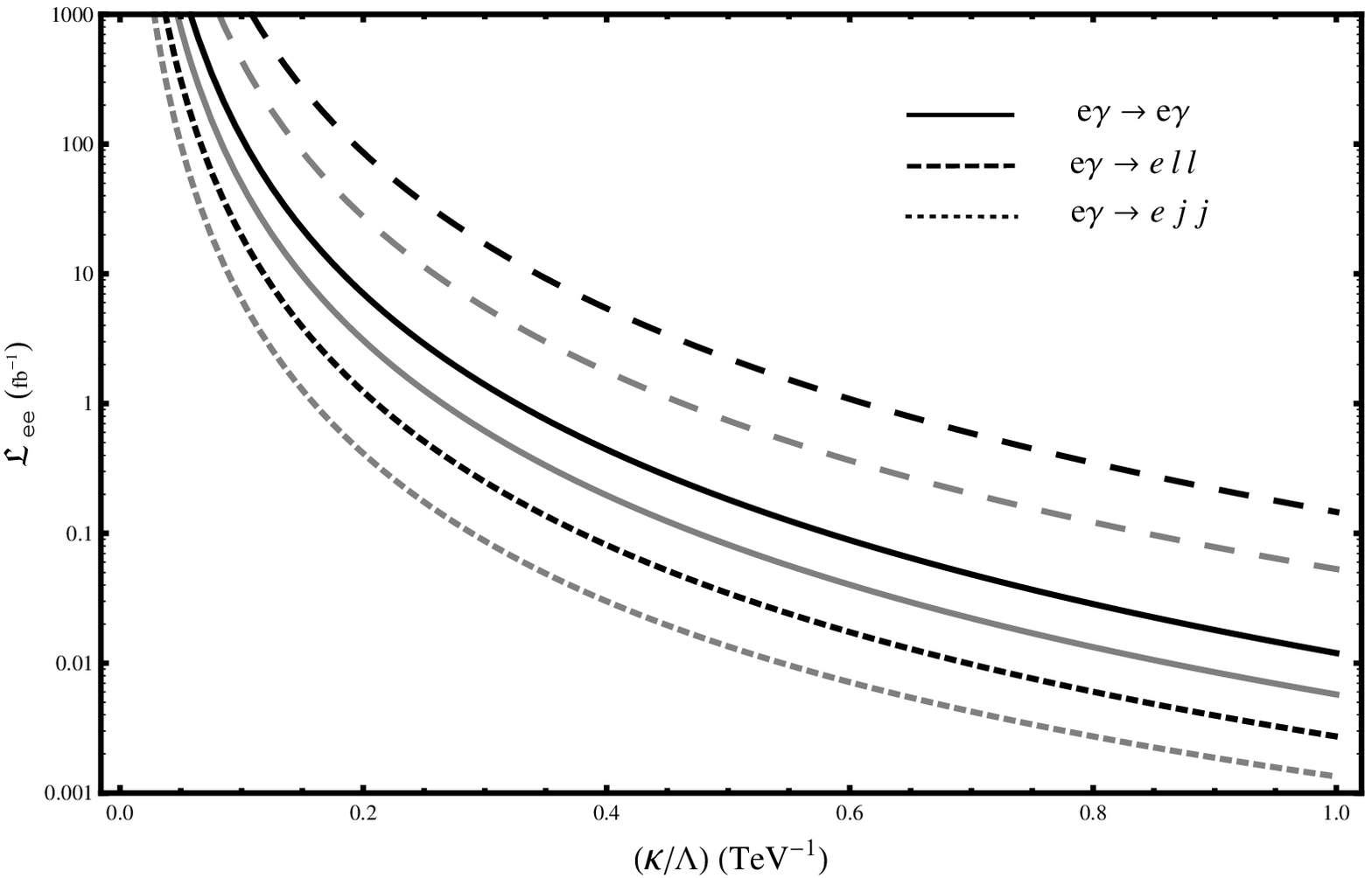}}
\vspace{-0.2cm}
\caption{The lowest necessary luminosity values of $e^+ e^-$ machines to observe anomalous processes at $\gamma e$ colliders (a) with $\sqrt{s}=500$ GeV and (b) with $\sqrt{s}=1$ TeV as a function of the charged lepton mass; (c) as a function of anomalous coupling strength for $m_{\ell_{4}}=300$ GeV at $\sqrt{s}=500$ GeV (black lines) and for $m_{\ell_{4}}=550$ GeV at $\sqrt{s}=1$ TeV  (grey lines). Cut selection 1 is used. \protect\label{fig6}}
\end{figure}

Achievable values of the anomalous coupling strengths are shown in Figs. 5a and 5b for $500$ GeV and $1$ TeV center of mass energies, respectively, as a function of the lepton mass. One can see that values as low as $0.027$ TeV$^{-1}$ ($0.023$ TeV$^{-1}$) are reachable for $\kappa/\Lambda$ with $\gamma e\rightarrow e j j$ process at $\sqrt{s}=500$ GeV with integrated luminosity of $100$ fb$^{-1}$ (at $\sqrt{s}=1$ TeV with $L_{int}=300$ fb$^{-1}$).  Moreover, it is possible to differ $\kappa_{\gamma}/\Lambda$ from $\kappa_{Z}/\Lambda$ by using informations coming from $\gamma e\rightarrow e \gamma$ or $\gamma e\rightarrow e Z$ processes, because corresponding cross sections are scaled as $(\kappa_{\gamma}/\Lambda)^2 \kappa_{\gamma}^2 /(\kappa_{\gamma}^2+\kappa_{Z}^2)$ and $(\kappa_{\gamma}/\Lambda)^2 \kappa_{Z}^2 /(\kappa_{\gamma}^2+\kappa_{Z}^2)$, respectively. The reachable values of anomalous photon and Z boson couplings are shown in Fig. 5c.

\begin{table}
\caption{Improvements with cut selection 2 for ideal $\sqrt{s}$.}  
\begin{ruledtabular}
\begin{tabular}{ccccc}
  & \multicolumn{2}{c}{$m_{\ell_{4}}=300$ GeV at $\sqrt{s}=331$ GeV }&\multicolumn{2}{c}{$m_{\ell_{4}}=550$ GeV at $\sqrt{s}=611$ GeV }   \\ 
 \cline{2-3} \cline{4-5}
  &Selection 1&Selection 2&Selection 1&Selection 2\\
\hline
$\gamma e\rightarrow \ell_{4} \rightarrow e\gamma $ & 0.049 & 0.025 & 0.032 & 0.019 \\
$\gamma e\rightarrow \ell_{4} \rightarrow e \ell \ell$ & 0.092 & 0.044 & 0.055 & 0.037 \\
$\gamma e\rightarrow \ell_{4} \rightarrow e j j$ & 0.026 & 0.018 & 0.019 & 0.014 \\
\end{tabular}
\end{ruledtabular}
\end{table}

The lowest necessary luminosities of $e^+ e^-$ machines to observe anomalous processes are plotted as a function of the lepton mass in Figs. 6a and 6b for $\sqrt{s}=500$ GeV and $1$ TeV options with $(\kappa/\Lambda)=1$ TeV$^{-1}$, respectively. It is seen that the single resonant production of the new charged lepton at $\gamma e$ colliders based on $e^+ e^-$ machines with $\sqrt{s}=500$ GeV will be observed almost in a working day for $m_{l_{4}}\geq 140$ GeV with $\gamma e\rightarrow e j j$ process. The $e \gamma \rightarrow e \gamma$ process becomes more advantageous at lepton masses smaller than $140$ GeV (Fig. 6a). A similar situation exists for the collider with $\sqrt{s}=1$ TeV (Fig. 6b). The Fig. 6c presents the lowest necessary luminosities as a function of the observation limit for the anomalous coupling strength for various cases. One can see that with $\sqrt{s}=1$ TeV and $300$ fb$^{-1}$ integrated luminosity $\kappa/\Lambda$ values down to $0.038$ TeV$^{-1}$ for $550$ GeV lepton mass will be reached. Similarly, the $\kappa/\Lambda$ values down to $0.066$ TeV$^{-1}$ for $m_{l_{4}}=300$ GeV can be observed with $\sqrt{s}=500$ GeV and $100$ fb$^{-1}$ integrated luminosity. The cut selection 2 causes improvements on achievable values of anomalous couplings. Table IV shows corresponding  improvements in the case of ideal $\sqrt{s}$ for $m_{l_{4}}=300$ and $550$ GeV.           

In conclusion, $\gamma e$ colliders will provide unique opportunity to search for anomalous couplings of the fourth SM family charged lepton. 

\begin{acknowledgments}

This work was supported by the Turkish Atomic Energy Authority (TAEA) and DPT with grant No. DPT2006K-120470.
 
\end{acknowledgments}

\end{document}